Sweet silver: a formaldehyde-free silver staining using aldoses as developing agents, with enhanced compatibility with mass spectrometry.

Mireille Chevallet [1,2], Sylvie Luche [1,2], Hélène Diemer [3], Jean-Marc Strub [3], Alain Van Dorsselaer [3], Thierry Rabilloud [1,2]

[1] CEA- Laboratoire de Biochimie et Biophysique des systèmes Intégrés, iRTSV/BBSI, CEA-Grenoble, 17 rue des martyrs, F-38054 GRENOBLE CEDEX 9, France

2 CNRS UMR 5092, CEA-Grenoble, 17 rue des martyrs, F-38054 GRENOBLE CEDEX 9, France

3 Laboratoire de Spectrométrie de Masse Bio-Organique, UMR CNRS 7178, ECPM, 25 rue Becquerel, 67087 STRASBOURG Cedex2, France

Correspondence :
Thierry Rabilloud, iRTSV/BBSI, UMR CNRS 5092,
CEA-Grenoble, 17 rue des martyrs,
F-38054 GRENOBLE CEDEX 9
Tel (33)-4-38-78-32-12
Fax (33)-4-38-78-44-99

e-mail: Thierry.Rabilloud@ cea.fr

Abstract

Protein detection methods after electrophoresis have to be sensitive, homogeneous, and not to impair downstream analysis of proteins by mass spectrometry. Speed, low cost and user-friendliness are also favored features. Silver staining combines many of these features, but its compatibility with mass spectrometry is limited. We describe here a new variant of silver staining that is completely formaldehyde-free. Reducing sugars in alkaline borate buffer are used as developers. While keeping the benefits of silver staining, this method is shown to afford a much better performance in terms of compatibility with silver staining, both in peptide mass fingerprinting by MALDI and in LC/ESI/MS/MS. .

1. Introduction

Protein detection is still a key step for the proteomics analysis of proteins separated on mono- or bidimensional gels. Besides obvious constraints such as sensitivity, homogeneity from one protein to another and linearity throughout a wide dynamic range, suitable features include compatibility with digestion and mass spectrometry, speed, convenience and low.

Up to now, no protein detection method matches perfectly these prerequisites, in the three families of methods that are of current use in proteomics.

Detection with organic dyes is simple, cheap and rather linear, shows an adequate compatibility with mass spectrometry, but its lacks sensitivity. In its most widely-used version, namely Colloidal Coomassie Blue [1], it requires long staining times for optimal sensitivity. However, faster protocols in this category are available, such as the dye pair staining technique [2], but the sensitivity remains moreless the same.

Fluorescent detection methods, on their side, show a convenient linearity and an adequate sensitivity, although the latter varies from the one of colloidal Coomassie, such as sypro orange [3] to the one of silver (sypro ruby, deep purple) [4], [5] . Although superior to the one of silver staining, their compatibility with mass spectrometry does not always equal the one of Coomassie Blue, and this has been unfortunately shown to be the case for Sypro Ruby [6] and Deep Purple [7], i.e. the most sensitive variants.

In addition to this drawback, optimal sensitivity requires rather long staining times, and the cost of the commercial reagents can become a concern when large series of gels are to be produced.

Within this category, the covalent labeling used in DIGE is slightly different [8], in the sense that the sensitivity in detection is reached by exploiting the very low fluorescent noise, thereby allowing to use the very high signal to noise ratio to achieve sensitive detection through a high-performance hardware. However, the absolute level of signal is very low in this technique, so that protein excision for identification by MS is made on a more heavily loaded gel stained with noncovalent fluorescent probes. Furthermore, spot excision after fluorescent staining is usually carried out on a UV table, with the associated safety problems.

The last family of protein detection methods consists of silver staining [9]. This method is sensitive, but labor-intensive, and its linearity is limited. However, in the proteomics frame, the most important problem lies in its limited compatibility with mass spectrometry. Although this feature has been improved by destaining of the spots or bands after silver staining [10], or by the use of silver-ammonia methods [11], the compatibility with mass spectrometry remains far below what can be achieved with fluorescent probes or colloidal Coomassie [12]. This low compatibility has been attributed to the use of formaldehyde [13], which also induces artefactual formylations. [14]. It would be therefore of great interest to have in hands a sensitive silver staining method totally formaldehyde-free. However, most silver reducers used in silver reduction (e.g. hydroquinone) [15] also show an important ability to induce protein crosslinks. To date, the only formaldehyde-free silver staining method uses carbohydrazide as the reducing agent [13]. While this results in improved compatibility with mass spectrometry, the staining performances are inadequate, both in sensitivity and homogeneity.

We describe here a new family of silver staining protocols, using reducing sugars in alkaline borate buffer as developing agents. These methods combine the classical sensitivity and staining homogeneity of classical silver staining methods, while showing a much improved compatibility with mass spectrometry.

2. Material and methods

2.1. Samples

Molecular weight markers (broad range, Bio-Rad) were diluted down to 10 ng/µl for each band in SDS buffer (Tris-HCl 125mM pH 7.5, containing 2% (w/v) SDS, 5% (v/v) thioglycerol, 20% (v/v) glycerol and 0.005% (w/v) bromophenol blue). The diluted solution was heated in boiling water for 5 minutes. A tenfold dilution in SDS buffer was performed to get a 1ng/µl per protein dilution

.

J774 cells (mouse macrophage) and WEHI274 cells (mouse monocytes) cells were grown in spinner flasks in DMEM + 5% fetal calf serum up to a density of 1 million cells /ml. The cells were collected by centrifugation (1000g 5 minutes), washed in isotonic wash buffer (10mM Tris HCl pH 7.5, 1mM EDTA and 250mM sucrose) and centrifuged at 100g for 5 minutes . The final pellet was suspended in its volume of isotonic wash buffer, transferred in an ultracentrifuge tube, and 4 volumes of concentrated lysis solution (8.75M urea, 2.5M thiourea, 5% CHAPS, 50mM DTT and 25mM spermine base) were added. After lysis at room temperature for 30 minutes, the viscous lystae was centrifuged at 200,000g for 1 hour at room temperature. The supernatant was collected, the protein concentration was estimated and the solution was made 0.4% (w/v) in carrier ampholytes (Pharmalyte 3-10). The solution was stored frozen at -20°C until use

2.2 Electrophoresis

2.2.1. SDS electrophoresis

10%T gels (160x200x1.5 mm) were used for protein separation. The Tris taurine buffer system was used [16], operated at a ionic strength of 0.1 and a pH of 7.9. The final gel composition is thus Tris 180mM, HCl 100 mM, acrylamide 10% (w/v), bisacrylamide 0.27%. The upper electrode buffer is Tris 50mM, Taurine 200mM, SDS 0.1%. The lower electrode buffer is Tris 50mM, glycine 200mM, SDS 0.1%.

For 1D SDS gels, a 4% stacking gel in Tris 125mM, HCl 100mM was used. No stacking gel was used for 2D electrophoresis.

The gels were run at 25V for 1hour, then 12.5W per gel until the dye front has reached the bottom of the gel.

2.2.2. IEF

Home made 160mm long 4-8 or 3-10.5 linear pH gradient gels were cast according to published procedures [17]. Four mm-wide strips were cut, and rehydrated overnight with the sample, diluted in a final volume of 0.6ml of rehydration solution (7M urea, 2M thiourea, 4% CHAPS and 100mM dithiodiethanol [18], [19]).

The strips were then placed in a multiphor plate, and IEF was carried out with the following electrical parameters

100V for 1 hour, then 300V for 3 hours, then 1000V for 1 hour, then 3400 V up to 60-70 kVh.

After IEF, the gels were equilibrated for 20 minutes in Tris 125mM, HCl 100mM, SDS 2.5%, glycerol 30% and urea 6M. They were then transferred on top of the SDS gels and sealed in place with 1% agarose dissolved in Tris 125mM, HCl 100mM, SDS 0.4% and 0.005% (w/v) bromophenol blue. Electrophoresis was carried out as described above.

2.3. Detection on gels

Colloidal coomassie blue staining was performed according to the published method [1]. Fluorescent staining was carried out with a ruthenium complex [6] with the improved protocol previously described [20].
The classical silver staining methods used were an ultrafast method [21], a silver-ammonia method [11] and a classical silver nitrate staining [22]

The new staining methods were based on the fast silver nitrate method, and all the steps up to silver impregnation were kept constant (see table 1). Only the developing bath was changed. Various reducing sugars (hexoses or pentoses) were tested at concentrations carrying from 15 to 150 mM. As development proceeds only under alkaline conditions, various alkaline buffers were tested, including sodium carbonate, sodium borate, sodium phosphate and sodium hydroxide, at pH ranging from 11 to 12.5.

The stop bath was the Tris-acetate buffer used in the silver nitrate method.

2.4. Image analysis
The gel images, acquired on an Agfa DuoScan T1200 at 300ppi resolution and grayscale mode, were converted to the TIFF format, and then analyzed with the delta 2D (v 3.5) software (Decodon, Germany). The default detection parameters calculated by the software were used and no manual edition of the spots was performed.

2.5. Mass spectrometry

2.5.1. Spot excision:

For fluorescent stain, spot excision was performed on a UV table operating at 302nm. The spots were collected in microtiter plates. The spots coming from gels stained with organic compounds (dyes or fluorophores) were not destained prior to acetonitrile washing. The spots coming from silver-stained gels were destained with the ferricyanide-thiosulfate protocol [10]. The solvent was then removed and the spots were stored at -20°C until use.

2.5.2. In gel digestion :

In gel digestion was performed with an automated protein digestion system, MassPrep Station (Waters Corp., Milford, USA). The gel plugs were washed twice with 50 µL of 25 mM ammonium hydrogen carbonate (NH4HCO3) and 50 µL of acetonitrile. The cysteine residue were reduced by 50 µL of 10 mM dithiothreitol at 57°C and alkylated by 50 µL of 55 mM iodoacetamide. After dehydration with acetonitrile, the proteins were cleaved in gel with 10 µL of 12.5 ng/µL of modified porcine trypsin (Promega, Madison, WI, USA) in 25 mM NH4HCO3. The digestion was performed overnight at room temperature. The generated peptides were extracted with 60% acetonitrile in 5% acid formic.

2.5.3. MALDI-TOF-MS analysis

MALDI-TOF mass measurements were carried out on UltraflexTM TOF/TOF (Bruker Daltonik GmbH, Bremen, Germany). This instrument was used at a maximum accelerating potential of 25kV in positive mode and was operated in reflectron mode. The samples were prepared by standard dried droplet preparation on stainless steel MALDI targets using alpha-cyano-4-hydroxycinnamic acid as matrix.

The external calibration of MALDI mass spectra was carried out using singly charged monoisotopic peaks of a mixture of bradykinin 1-7 (m/z=757.400), human angiotensin II (m/z=1046.542), human angiotensin I (m/z=1296.685), substance P (m/z=1347.735), bombesin

(m/z=1619.822), renin (m/z=1758.933), ACTH 1-17 (m/z=2093.087) and ACTH 18-39 (m/z=2465.199). To achieve mass accuracy, internal calibration was performed with tryptic peptides coming from autolysis of trypsin, with respectively monoisotopic masses at m/z = 842.510, m/z = 1045.564 and m/z = 2211.105. Monoisotopic peptide masses were automatically annotated using Flexanalysis 2.4 software. Peaks are automatically collected with a signal to noise ratio above 4 and a peak quality index greater than 30.

2.5.4 LC-MS/MS

Nano-LC-MS/MS analysis was performed either using a nanoAcquity UPLC$^{TM}$ system (Waters Corp., Milford, USA), coupled to a Synapt HDMS$^{TM}$ mass spectrometer (Waters Corp., Milford, USA).

From each sample, 4.5 µL was loaded on a precolumn (Waters, C18, 5µm, 180 µm id, 20 mm length), before chromatographic separation on a C18 column (Waters, C18, 1.7 µm, 75 mm id, 200 mm length). The gradient was generated at a flow rate of 400 nL/min. The gradient profile consisted of a linear one from 99% of a water solution acidified by 0.1% HCOOH vol/vol (solution A), to 50% of a solution of CH3CN acidified by 0.1% HCOOH vol/vol (solution B) in 35 min, followed by a second gradient ramp to 90% of B in 1 min. Data acquisition was piloted by MassLynx software V4.1. Calibration was performed using adducts of 0.1% phosphoric acid (Acros, NJ, USA) with a scan range from m/z 50 to 2000. Automatic switching between MS and MS/MS modes was used. The internal parameters of the Synapt HDMS$^{TM}$ were set as follows. The electrospray capillary voltage was set to 3.2 kV, the cone voltage set to 35 V, and the source temperature set to 80°C. The MS survey scan was m/z 250–1500 with a scan time of 1 s. When the peak intensity of 2+, 3+ or 4+ peptide ions rose above a threshold of 12 counts/s, tandem mass spectra were acquired. The scan range for MS/MS acquisition was from m/z 50 to 2000 with a scan time of 1. Fragmentation was performed using argon as the collision gas and with a collision energy profile optimized for various mass ranges of precursor ions. Data processing was done automatically with the ProteinLynx Global server V.2.3 (Waters Corp., Milford, USA).

2.5.5. MS and MS/MS Data analysis

The MASCOT search algorithm (Version 2.2.04, Matrix Science, London, UK) [23] was used for protein identification against the Swiss-Prot database (55.1). All proteins present in the database were used without any pI and Mr restrictions. A maximum number of one missed cleavage by trypsin was allowed, and carbamidomethylated cysteine and oxidized methionine were set as variable modifications. For the peptide mass fingerprint, the peptide mass error was limited to 50 ppm. For MS/MS ion search, only doubly and triply peptides were searched. The peptide tolerance was typically set to 50 ppm and MS/MS tolerance was set to 0.1 Da.

3. Results and discussion

The first step of this study was to build an efficient stain. The basic requisite was to eliminate completely the formaldehyde, and to replace it in the stain development step by another reducer. However, it was known from prior work that inorganic reducers [24] and many organic ones [13] do not lead to any practical staining. Thus, the choice was restricted to aldehydes However, despite the use of glutaraldehyde in some silver staining protocols [25], this class of chemicals have not been extensively tested. While bifunctional aldehydes were ruled out from their well-known protein crosslinking behavior, most aliphatic aldehydes either did not give any practical stain or were not soluble enough in water. The only exception to these rules seemed to be aldoses,, i.e. hydroxy aldehydes, which are fairly soluble in water and have a chemical reactivity which is different from the one of aliphatic aldehydes. The use of glucose has been metioned briefly in the literature [26], but this idea had to be brought to a stage usable in proteomics. Initial tests made by SDS electrophoresis of molecular weight markers showed us that a practical staining was obtained only if both the concentration of the sugar was in the 100mM range (2% w/v), which is almost 10 times the concentration of formaldehyde in a classical silver stain developer, and if at the same time the pH of the developer was increased to 12 or over. As carbonate solutions do not reach easily such values, we tested other basic solutions, namely sodium hydroxide, and phosphate or borate buffer. As shown on figure 1, a weak staining was obtained with sodium hydroxide (center panel) while phosphate gave absolutely no stain and borate gave the best result, probably because of its well-known binding to the sugar diols structures.

Further corroborating this view of a specific buffer effect, we could not devise any stain using both silver ammonia as the silvering agent and a sugar in the developing agent (data not shown).

This borate-specific effect led us to investigate if some sugars would perform better than others in such a protocol. To this purpose, we tested several hexoses and pentoses. We also increased the pH of the developer up to 12.7 to increase development speed and sensitivity, and the results are shown on figure 2. Except mannose, which did not give any staining (data not shown) all the other aldoses tested gave a positive staining. On the whole, pentoses were more efficient than hexoses, but this may be linked to their greater molar concentration at equal weight. This led us to the final protocol for formaldehyde free silver staining, shown on table 1.

As this stain is intended to be used for proteomics analysis, its reproducibility needed to be verified. To this purpose, an identical sample (200µg of J774 total cell extract was loaded on 6 different 2D gels (immobilized pH gradient pH 4-8). Three were stained by the classical silver staining, and three by the new silver-galactose-borate stain. The resulting images were then analyzed by the Delta 2D software. Approximately 1500 spots were detected on the gels (shown on supplemental figure 1) , independently of the stain used, indicating at the same time i) a close sensitivity between the new stain and the control one, and ii) a good reproducibility of each staining method. The relative standard deviation (i.e. the standard deviation/ mean volume, expressed in percentile) was calculated for each spot. The median rsd (i.e. the value for which half of the spots have a lower rsd and half a higher one) was 16% for the control silver staining and 11% for the silver-galactose-borate stain. This comparable median rsd further documents the reproducibility of the new stain as at least as good as the one of the control stain.

Finally, the practicability of the whole method was assessed by comparison between three variants of the silver-aldose staining with three classical variants of silver staining (ultrafast, standard silver nitrate and silver ammonia) as well as with colloidal Coomassie staining and fluorescent staining (Figure 3) . Homologous spots were excised on each gel and analyzed by mass spectrometry. The results are shown on table 2, from which several trends can be drawn::

(i) as expected, the overall compatibility with mass spectrometry of formaldehyde-free silver staining methods is higher than the one of classical ones, especially in the high molecular weight range.

(ii) among the aldose-silver staining variants, galactose shows the best compromise between staining performances and performances in subsequent MS analysis

(iii) the high performance silver staining methods, and especially the improved ones, allow to visualize and analyze spots that escape detection by colloidal Coomassie or fluorescence and even by medium sensitivity silver staining [21] (e.g. vinculin, plastin or importin).

These results can be seen directly on the mass spectra, as shown on figure 4.

We also checked that the better performances of the sugar developers in terms of mass spectrometry compatibility were not restricted to MALDI mass spectrometry. To this purpose, two spots (vinculin and malate dehydrogenase) were analyzed by LC-ESI/MS/MS, and the summarized results are reported on Table 3. It can be seen that the better MS compatibility of sugar developers applies also with this mass spectrometry method, so that the improvement over formaldehyde developer is likely to take place at the digestion/peptide extraction step.

Finally, we checked the overall performance of the new stain on a wide pH range and at a lower protein load. The results, shown on figure 5, demonstrate that the silver-aldose stain is slightly less sensitive than the classical silver staining, but that there is no gross difference over the complete pH range between the two stains.

Although the better compatibility with silver staining undoubtedly supports a better yield of unmodified peptides, as those ones only are counted positive by our criteria, it cannot be ruled out that the aldoses, being aldehydes, can modify some reactive groups in the proteins, especially the side chain amino group of lysines. However, the missed cleavage at this site, plus the mass of the glucide, are likely to produce heavy peptides that will not easily show up in mass spectrometry. However, if such peptides are detected, this artefactual modification could be mistaken for a glycation, a modification associated with aging [27], [28]. However, glycation has been described to date only with glucose. This is why we tested pentoses as developing agents, as these sugars will induce a completely unnatural modification. Moreover, if a natural modification of the same mass is expected, performing a duplicate experiment where one gel is developed with a hexose and the other gel with a pentose would discriminate between natural and artefactual modifications.

On the point of view of peptide extraction, the combination of a very short fixation, as in the Shevchenko's method, with this sugar-borate developer would probably further enhance peptide recovery. However, for reasons that remain unclear, we could not make a practically usable stain combining both features.

4. Concluding remarks

We believe that the sugar-borate developer brings silver staining to a happy compromise for proteomics. It keeps the advantages of visible methods, namely the absence of requirement of costly hardware (required for fluorescence). it affords a better sensitivity than colloidal Coomassie, fluorescence and silver staining methods really optimized for downstream mass spectrometry (e.g. the Shevchenko's method). Although the sensitivity of this silver staining method is slightly inferior to the one of the best silver staining methods, this is outweighed by the superior mass spectrometry compatibility, and by the absence of fomaldehyde-linked artifacts. It also allows for a better safety in the laboratory, as this method does not require either a UV table for spot excision (required by many fluorescent methods) nor the noxious formaldehyde chemical.

5. Acknowledgments.

The support from the Région Rhone Alpes by a grant (analytical chemistry subcall, priority research fields call 2003-2006) is gratefully acknowledged.

6. References

[1] Neuhoff, V., Arold, N., Taube, D., Ehrhardt, W. Improved staining of proteins in polyacrylamide gels including isoelectric focusing gels with clear background at nanogram sensitivity using Coomassie Brilliant Blue G-250 and R-250. Electrophoresis 1988, 9, 255-262

[2] Choi, J.K., Tak, K.H., Jin, L.T., Hwang, S.Y., et al. Background-free, fast protein staining in sodium dodecyl sulfate polyacrylamide gel using counterion dyes, zincon and ethyl violet. Electrophoresis. 2002, 23, 4053-4059

[3] Steinberg, T.H., Jones, L.J., Haugland, R.P., Singer, V.L. SYPRO orange and SYPRO red protein gel stains: one-step fluorescent staining of denaturing gels for detection of nanogram levels of protein. Anal Biochem. 1996, 239, 223-237.

[4] Berggren, K., Chernokalskaya, E., Steinberg, T.H., Kemper, C. Background-free, high sensitivity staining of proteins in one- and two-dimensional sodium dodecyl sulfate-polyacrylamide gels using a luminescent ruthenium complex. Electrophoresis. 2000, 21, 2509-2521.

[5] Mackintosh, J.A., Choi, H.Y., Bae, S.H., Veal, D.A., et al. A fluorescent natural product for ultra sensitive detection of proteins in one-dimensional and two-dimensional gel electrophoresis. Proteomics. 2003, 3, 2273-2288

[6] Rabilloud, T., Strub, J.M., Luche, S., van Dorsselaer, A., Lunardi, J. A comparison between Sypro Ruby and ruthenium II tris (bathophenanthroline disulfonate) as fluorescent stains for protein detection in gels．Proteomics. 2001, 1, 699-704.

[7] Chevalier, F., Centeno, D., Rofidal, V., Tauzin, M. et al. Different impact of staining procedures using visible stains and fluorescent dyes for large-scale investigation of proteomes by MALDI-TOF mass spectrometry. J Proteome Res. 2006, 5, 512-520

[8] Unlu, M., Morgan, M.E., Minden, J.S. Difference gel electrophoresis: a single gel method for detecting changes in protein extracts. Electrophoresis. 1997, 18, 2071-2077

[9] Chevallet, M., Luche, S., Rabilloud, T. Silver staining of proteins in polyacrylamide gels. Nat Protoc. 2006,1, 1852-1858

[10] Gharahdaghi, F., Weinberg, C.R., Meagher, D.A., Imai, B.S., Mische, S.M. Mass spectrometric identification of proteins from silver-stained polyacrylamide gel: a method for the removal of silver ions to enhance sensitivity. Electrophoresis. 1999, 20, 601-605.

[11] Chevallet, M., Diemer, H., Luche, S., van Dorsselaer, A. et al. Improved mass spectrometry compatibility is afforded by ammoniacal silver staining. Proteomics 2006. 6, 2350-2354.

[12] Winkler, C., Denker, K., Wortelkamp, S., Sickmann, A.. Silver- and Coomassie-staining protocols: detection limits and compatibility with ESI MS. Electrophoresis. 2007, 28, 2095-2099

[13] Richert, S. Luche S, Chevallet M, Van Dorsselaer A. About the mechanism of interference of silver staining with peptide mass spectrometry. Proteomics. 2004, 4, 909-916

[14] Osés-Prieto, J.A., Zhang, X., Burlingame, A.L. Formation of epsilon-formyllysine on silver-stained proteins: implications for assignment of isobaric dimethylation sites by tandem mass spectrometry. Mol Cell Proteomics. 2007, 6, 181-92

[15] Merril, C.R. Silver stains for proteins in gels : US patent 4,405,720 1983

[16] Tastet, C., Lescuyer, P., Diemer, H., Luche, S., et al. .A versatile electrophoresis system for the analysis of high- and low-molecular-weight proteins.Electrophoresis. 2003, 24, 1787-1794.

[17] Rabilloud, T., Valette, C., Lawrence, J.J. Sample application by in-gel rehydration improves the resolution of two-dimensional electrophoresis with immobilized pH gradients in the first dimension . Electrophoresis, 1994, 15, 1552-1558

[18] Rabilloud, T., Adessi, C., Giraudel, A., Lunardi, J., Improvement of the solubilization of proteins in two-dimensional electrophoresis with immobilized pH gradients. Electrophoresis, 1997, 18, 307-316

[19] Luche, S., Diemer, H., Tastet, C., Chevallet, M., et al. About thiol derivatization and resolution of basic proteins in two-dimensional electrophoresis. Proteomics. 2004, 4, 551-561.

[20] Lamanda, A., Zahn, A., Roder, D., Langen, H. Improved Ruthenium II tris (bathophenantroline disulfonate) staining and destaining protocol for a better signal-to-background ratio and improved baseline resolution. Proteomics. 2004, 4, 599-608

[21] Shevchenko, A., Wilm, M., Vorm, O., Mann, M. Mass spectrometric sequencing of proteins silver-stained polyacrylamide gels. Anal Chem. 1996, 68, 850-858.

[22] Rabilloud, T. A comparison between low background silver diammine and silver nitrate protein stains. Electrophoresis. 1992, 13, 429-439

[23] Perkins, D.N., Pappin, D.J., Creasy, D.M., Cottrell, J.S. Probability-based protein identification by searching sequence databases using mass spectrometry data., Electrophoresis 1999; 20, 3551-3567

[24] Heukeshoven, J, Dernick, R. Simplified method for silver staining of proteins in polyacrylamide gels and the mechanism of silver staining. Electrophoresis 1985, 6, 103-112.

[25] Porro, M., Viti, S., Antoni, G., Saletti, M. Ultrasensitive silver-stain method for the detection of proteins in polyacrylamide gels and immunoprecipitates on agarose gels. Anal Biochem. 1982 127, 316-321

[26] Heukeshoven, J., Dernick, R., Neue Ergebnisse der Silberfaerbung von proteinen : Untersuchungen zum Mechanismus und Einfluss einiger Parameter auf die Effektivitaet der Silberfaerbung. In: Radola, B.J. (Ed.), Electrophorese Forum' 85, Technische Universitaet Muenchen 1985, pp. 108-118.

[27] Suji, G., Sivakami, S. Glucose, glycation and aging. Biogerontology. 2004, 5, 365-373.

[28] Ulrich, P., Cerami, A. Protein glycation, diabetes, and aging. Recent Prog Horm Res. 2001, 56, 1-21.

Legends to figures

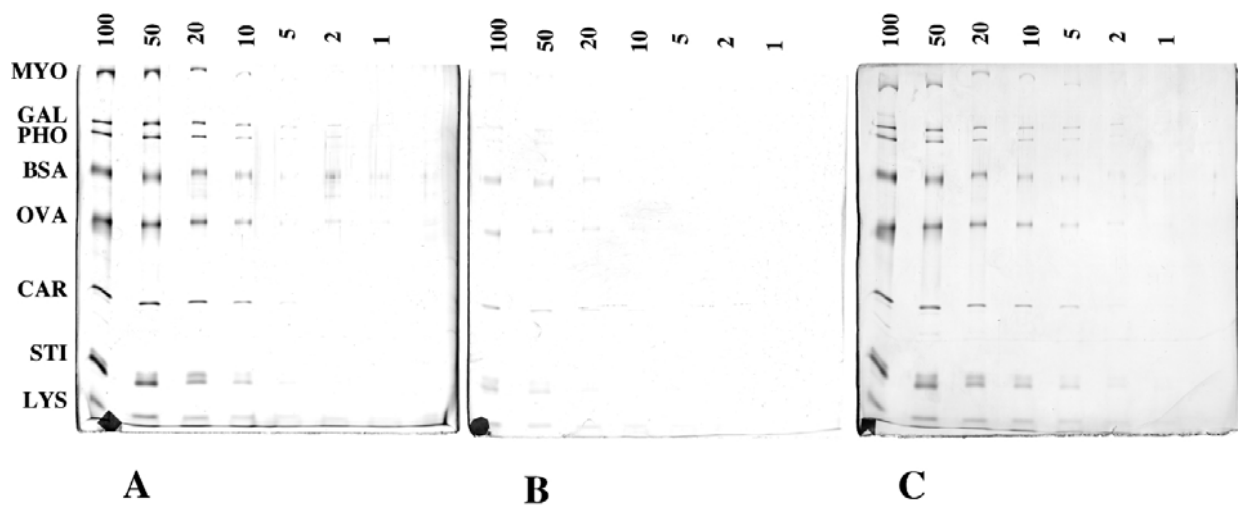

Figure 1: Initial tests of silver staining with aldoses

Molecular weight markers (BioRad, broad range) were diluted serially in SDS buffer and separated by SDS electrophoresis. The corresponding gel was stained with standard (formaldehyde) silver staining (panel A), or with glucose as the developing agent in 3.5% potassium carbonate (panel B) or in 100mM boric acid/150mM NaOH (panel C). Protein separated: MYO: myosin (205kDa), GAL: beta galactosidase (116kDa), PHO: glycogen phosphorylase (97kDa) BSA: bovine serum albumin (67kDa), OVA: ovalbumin (46 kDa), CAR: carbonic anhydrase (30 kDa), STI: soybean trypsin inhibitor (21 kDa), LYS: lysozyme (14.5 kDa). Protein content per lane in each panel, from left to right: 100 ng/protein, 100 ng, 50 ng, 20 ng, 10 ng, 5 ng, 2 ng. 1 ng

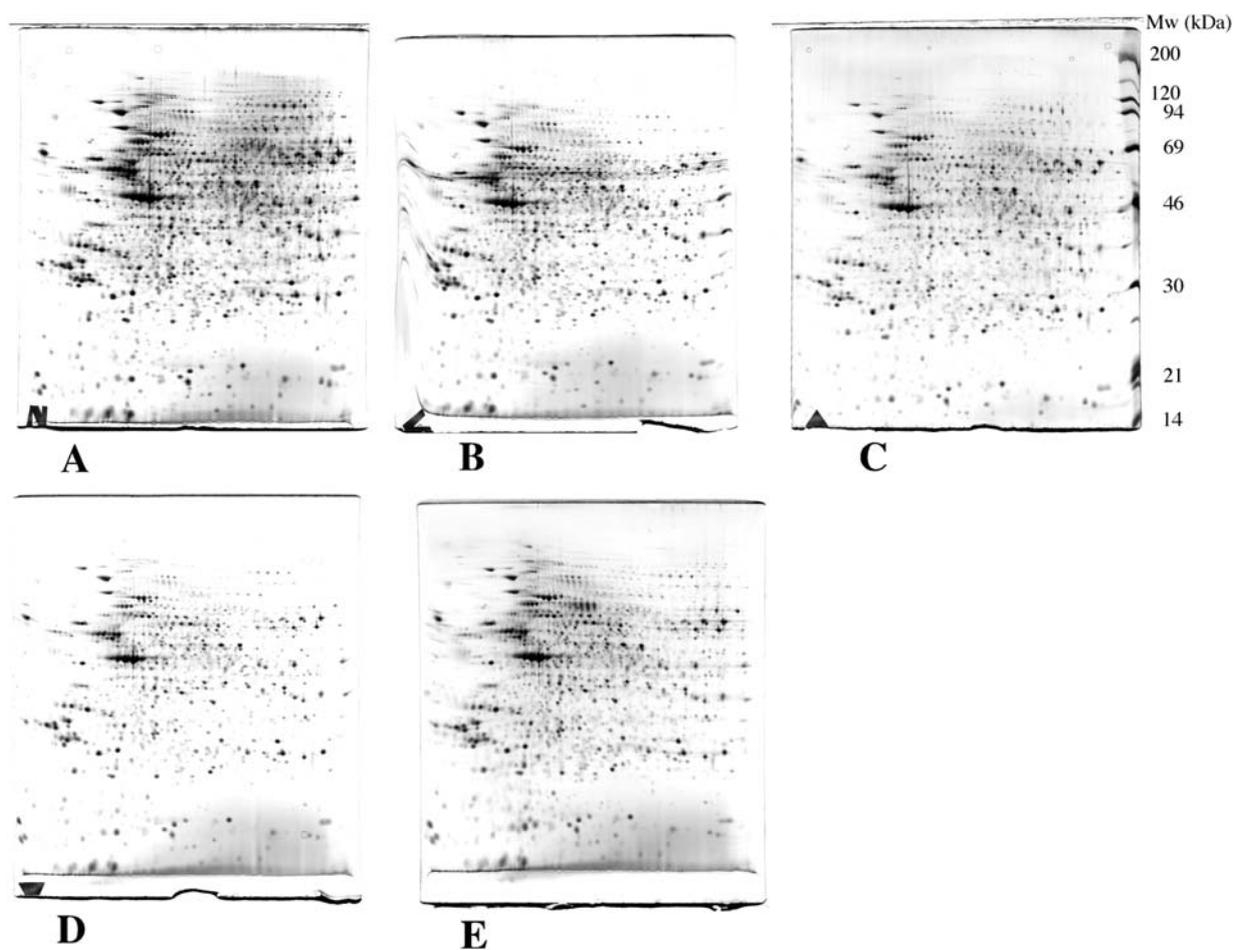

Figure 2 Comparison of various aldoses as developing agents

200μg of proteins extracted from WEHI 274 cells were separated by 2D gel electrophoresis and detected by various methods. The IEF pH range was 4-8 linear, and the xecond dimension gel was a 10% continuous gel. The gels were stained by silver nitrate, with 2% aldose in 100mM borate buffer pH 12.7 in developer

A: arabinose. B: ribose. C: xylose. D: galactose E: glucose

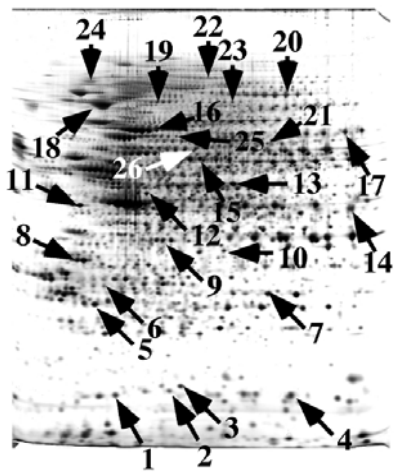
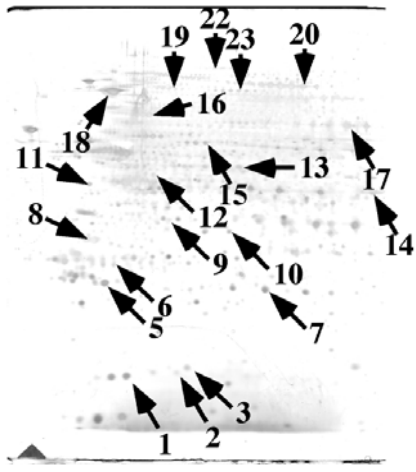
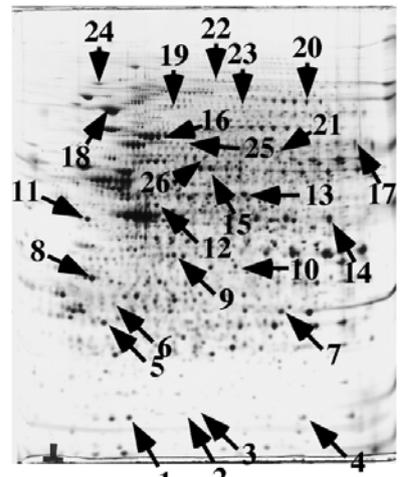
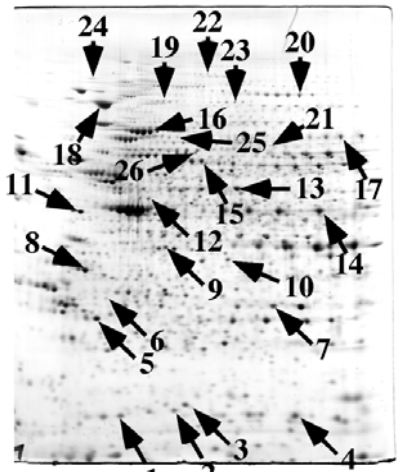
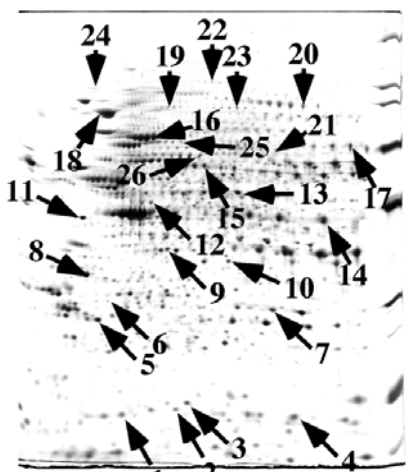
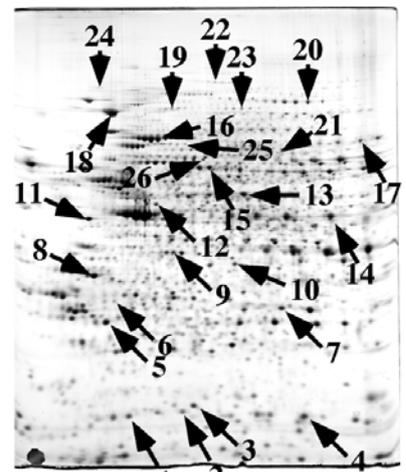
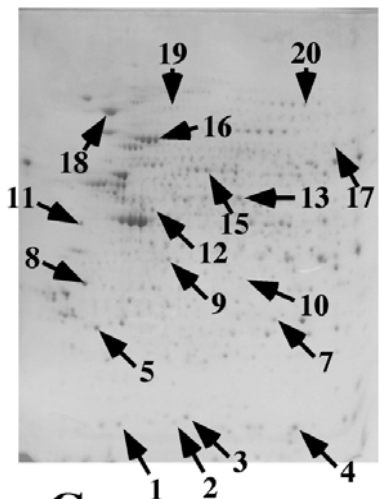
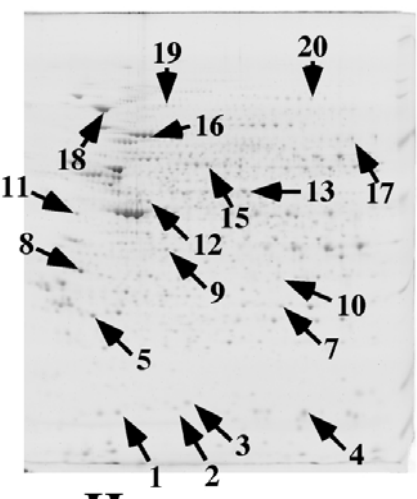

Figure 3: comparison of various staining methods

200µg of proteins extracted from J774 murine macrophages were separated by 2D gel electrophoresis and detected by various methods. The IEF pH range was 4-8 linear, and the xecond dimension gel was a 10% continuous gel.

A: detection by silver staining, formaldehyde developer. B: detection by ultrafast silver staining (Shevchenko's method). C: detection by silver ammonia. D: detection by silver-glucose-borate. E: detection by silver-galactose-borate. F: detection by silver-xylose-borate. G: detection by colloidal coomassie blue. H: detection by fluorescence (ruthenium complex). The spots excised for mass spectrometry analysis are shown by arrows

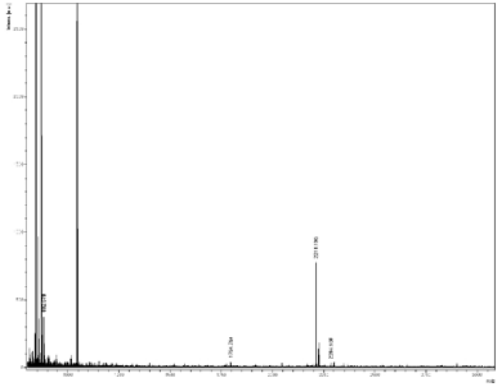
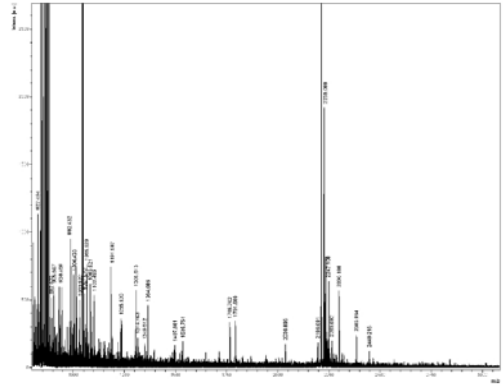
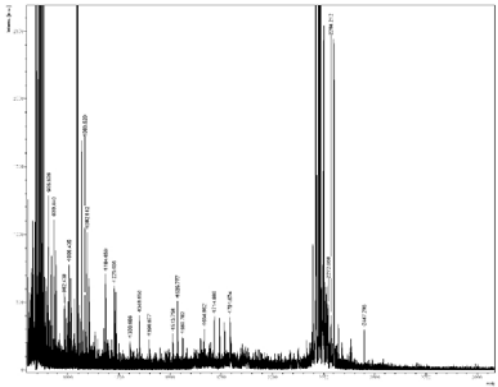
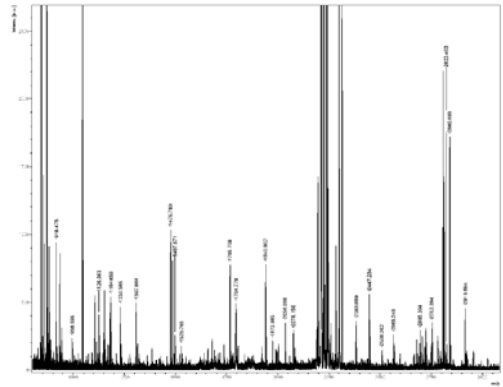
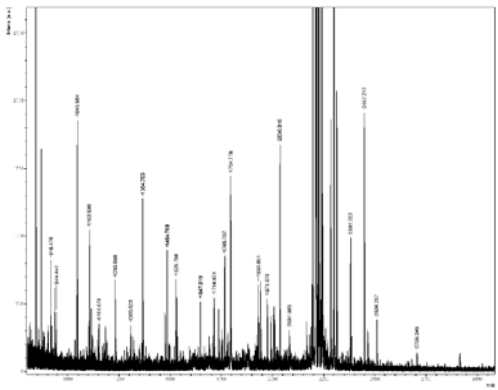
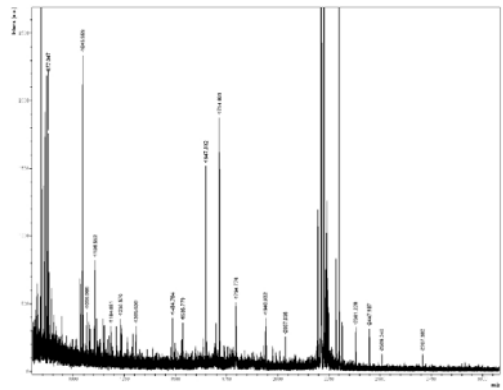

Figure 4: MALDI spectra comparison

The MALDI spectra coming from the vinculin spots excised from the gels shown on figure 5 are compared. The spectra are displayed at the same scale, and were recorded under the same apparatus conditions (e.g. number of laser shots, laser fluence etc…). The spectra came from spts stained with the different silver staining methods

A: silver ammonia; B: silver nitrate; C: ultrafast silver; D: silver glucose borate; E: silver galactose borate; F: silver xylose borate.

The better extraction of peptides above 1500 Da, containing more sequence information, is obvious in this case with the silver-aldose methods 75µg of proteins extracted from J774 murine macrophages were separated by 2D gel electrophoresis and detected by various methods. The IEF pH range was 3.5-10.5 linear, and the xecond dimension gel was a 10% continuous gel.

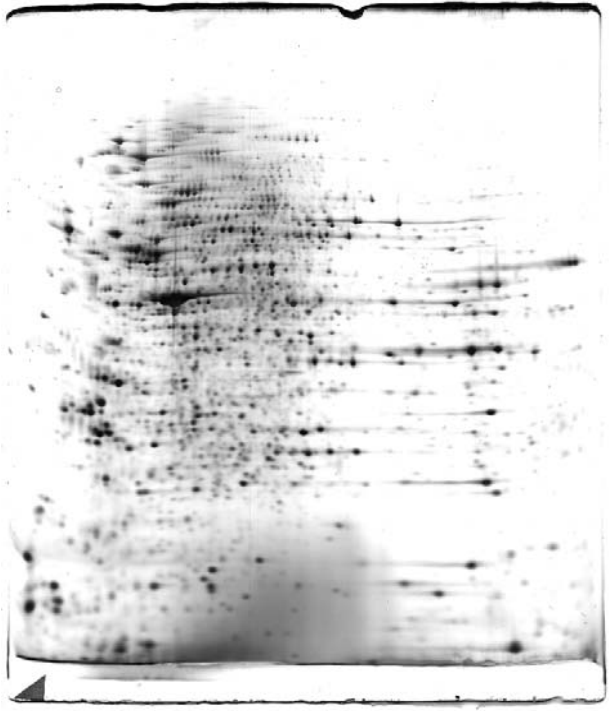 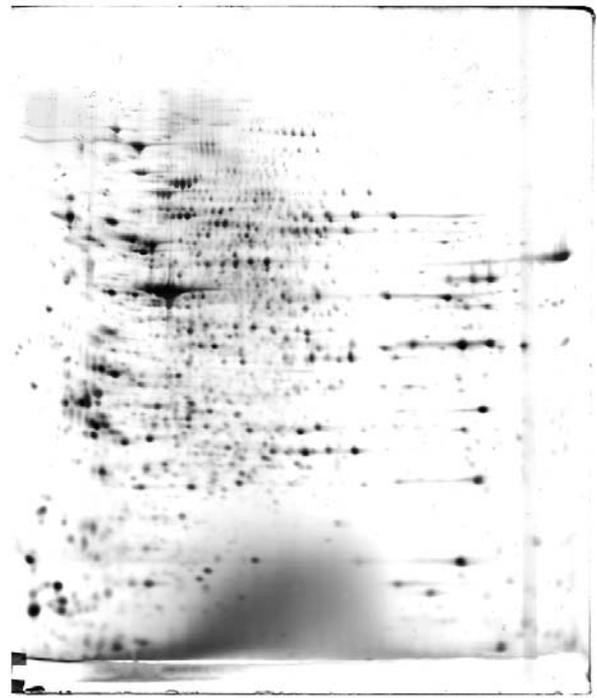

Figure 5: Evaluation of staining performances

A: detection by silver staining, formaldehyde developer. B: detection by silver-galactose-borate.

Table 1: flowchart for silver staining

| Step | solution | Time |
|---|---|---|
| Fixation | Ethanol 30% (v/v) Acetic acid 10% (v/v) | overnight |
| Rinse | water | 4x 10 minutes |
| Sensitization | 8mM sodium thiosulfate | 1 minute |
| Rinse | water | 2x 1 minute |
| Silvering | Silver nitrate 12mM | 20-30 minutes |
| Rinse | water | 5-10 seconds |
| development | 2% (w/v) sugar, 100mM boric acid, 150mM NaOH, 50µM sodium thiosulfate | 20-30 minutes |
| Stop | Tris 40g/l, acetic acid 20 ml/l | 30 minutes |

This table represents the optimized protocol for silver staining with aldoses. In the initial siolver staining, the developer contains 3.5% potassium carbonate, 50µM thiosulfate and 3mM formaldehyde

| spot | protein | Accession number | CCB | Rubps | Ag control | Ultrafast | Ag ammonia | Ag glucose | Ag galactose | Ag xylose |
|---|---|---|---|---|---|---|---|---|---|---|
| 1 | eIF5A | P63242 | 83/152 | 51/75 | NI | 68/81 | 63/134 | 62/89 | 79/74 | 46/78 |
| 2 | ARP 5 | Q9CPW4 | ND | ND | 69/119 | 76/112 | 61/73 | NI | 45/54 | 45/56 |
| 3 | stahmin | P54227 | 55/118 | 51/96 | 59/176 | 64/131 | 72/177 | 25/57 | 35/77 | 36/72 |
| 4 | PCTI | P17742 | 58/174 | 54/130 | 51/150 | ND | 62/155 | 46/132 | 67/114 | 63/111 |
| 5 | Rho GDI-2 | Q61599 | ND | 32/55 | 51/65 | 61/147 | 61/174 | 48/96 | 55/107 | 66/104 |
| 6 | IPA | O55023 | ND | 26/194 | NI | 22/64 | NI | NI | 19/46 | NI |
| 7 | PGM | Q9DBJ1 | 53/186 | 21/37 | 48/131 | 51/128 | 24/60 | 58/193 | 68/190 | 58/194 |
| 8 | Annexin V | P48036 | 74/316 | 75/288 | 58/244 | 78/303 | 52/219 | 68/205 | 72/226 | 73/266 |
| 9 | Annexin III | O35639 | 75/348 | NI | 55/258 | 64/289 | 55/181 | 53/167 | 50/197 | NI |
| 10 | MDH | P14152 | 42/134 | NI | NI | 35/73 | NI | 35/108 | 33/76 | 23/63 |
| 11 | RPSA | P14206 | 50/154 | 28/76 | 36/158 | 41/170 | 41/169 | 43/119 | 43/111 | 52/120 |
| 12 | eIF 4A | P60843 | 53/315 | 46/192 | 51/235 | NI | 49/198 | 54/228 | 60/255 | NI |
| 13 | enolase | P17182 | 72/355 | 54/194 | 42/153 | 70/330 | 55/124 | 58/231 | 52/201 | 57/238 |
| 14 | PGK | P09411 | 67/301 | NI | 55/182 | 55/180 | 55/221 | 52/158 | 62/202 | 52/155 |
| 15 | PDI | P27773 | 60/385 | 44/176 | 61/272 | 54/352 | 62/410 | 54/302 | 54/291 | 54/298 |
| 16 | HSC 71 | Q3U9G0 | 52/221 | 47/161 | 56/253 | 48/260 | 51/180 | 53/181 | 55/197 | 49/193 |
| 17 | transketolase | P40142 | NI | 47/137 | 37/258 | 29/224 | NI | 41/175 | 39/183 | 31/151 |
| 18 | HSP90 | P11499 | 44/255 | 22/126 | 51/254 | 53/319 | 46/297 | 39/234 | 36/207 | 38/239 |
| 19 | gelsolin | P13020 | 14/80 | NI | NI | 28/237 | 28/119 | 29/154 | 23/90 | 23/90 |
| 20 | EF2 | P58252 | ND | 29/127 | 40/223 | 50/286 | 28/184 | 40/221 | 43/238 | 42/237 |
| 21 | lamin A/C | Q9DC21 | ND | ND | 48/275 | ND | 45/370 | NI | 20/109 | 26/140 |
| 22 | vinculin | Q64727 | ND | ND | 11/64 | 17/57 | NI | 25/132 | 23/106 | NI |
| 23 | LHCR2 | Q99KC8 | ND | ND | NI | 25/108 | NI | NI | 18/95 | 26/123 |
| 24 | importin | Q8BKC5 | ND | ND | 11/51 | ND | NI | 33/156 | 18/92 | NI |
| 25 | plastin-3 | Q99K51 | ND | ND | 17/57 | ND | NI | 14/48 | 35/159 | 54/155 |
| 26 | TCP1 epsilon | P11983 | ND | ND | NI | ND | NI | 18/59 | 19/68 | 36/128 |
| proportion of identified spots | | | 15/26 | 15/26 | 20/26 | 20/26 | 18/26 | 22/26 | 26/26 | 21/26 |

Table 2: MS analysis of proteins stained by different methods

Homologous spots excised from two-dimensional gels (4-8 linear pH gradients, 10% acrylamide) loaded with equal amounts of J774 proteins (200µg) and stained by various methods were digested, and the digests were analysed by MALDI mass spectrometry. The summary of the mass spectrometry data is in the form %coverage / Mascot score. The spot numbers refer to those shown on figure 3.

CCB: colloidal -Coomassie blue staining. RuBPS: fluorescent staining with a ruthenium complex

NI: not identified. ND: not detected

Protein names: EF2: elongation factor 2; eIF: eukaryotic initiation factor; IPA: phosphoinositol phosphatase; LHCR2: Loss of heterozygosity 11 chromosomal region 2 gene A protein homolog ; MDH: malate dehydrohenase; PCTI: prolyl cis-trans isomerase; PDI: protein disulfide isomerase; PGK: phosphoglycerate kinase; PGM: phosphoglucomutase; RPSA: Ribosomal protein 40S subunit, protein SA;

| Vinculin | Q64727 | | |
|---|---|---|---|
| staining method | % coverage | nb of peptides | Mascot score |
| silver nitrate | 5 | 4 | 95 |
| silver ammonia | 2 | 2 | 63 |
| ultrafast silver | 4 | 4 | 208 |
| coomassie blue | ND | ND | ND |
| silver glucose | 15 | 13 | 442 |
| silver galactose | 19 | 15 | 485 |
| silver xylose | 9 | 8 | 177 |
| malate dehydrogenase | P14152 | | |
| silver nitrate | 14 | 4 | 144 |
| silver ammonia | 2 | 1 | 73 |
| ultrafast silver | 20 | 4 | 307 |
| coomassie blue | 27 | 7 | 283 |
| silver glucose | 27 | 8 | 361 |
| silver galactose | 23 | 8 | 282 |
| silver xylose | 29 | 9 | 385 |

Table 3: summary of MS/MS results obtained on vinculin and malate dehydrogenase